\documentclass[a4paper,draftcls,onecolumn,technote,11pt]{IEEEtran}%
\usepackage{amsfonts}
\usepackage{amsmath}
\usepackage{amssymb}
\usepackage{graphicx}%
\providecommand{\U}[1]{\protect\rule{.1in}{.1in}}
\begin{document}

\title{Controllability Analysis and Degraded Control for a Class of Hexacopters
Subject to Rotor Failures}
\author{Guang-Xun Du, Quan Quan, Kai-Yuan Cai\thanks{The authors are with Department
of Automatic Control, Beihang University, Beijing 100191, China
(dgx@asee.buaa.edu.cn; qq\_buaa@buaa.edu.cn; kycai@buaa.edu.cn)}}
\maketitle

\begin{abstract}
This paper considers the controllability analysis and fault tolerant control
problem for a class of hexacopters. It is shown that the considered hexacopter
is uncontrollable when one rotor fails, even though the hexacopter is
over-actuated and its controllability matrix is row full rank. According to
this, a fault tolerant control strategy is proposed to control a degraded
system, where the yaw states of the considered hexacopter are ignored.
Theoretical analysis indicates that the degraded system is controllable if and
only if the maximum lift of each rotor is greater than a certain value. The
simulation and experiment results on a prototype hexacopter show the
feasibility of our controllability analysis and degraded control strategy.

\end{abstract}


\section*{Nomenclature}%

\begin{tabular}
[c]{@{}lcl}%
$h$ & = & altitude of the multirotor helicopter, m\\
$\phi,\theta,\psi$ & = & roll, pitch and yaw angles of the multirotor
helicopter, rad\\
$v_{h}$ & = & vertical velocity of the multirotor helicopter, m/s\\
$p,q,r$ & = & roll, pitch and yaw angular velocities of the multirotor
helicopter, rad/s\\
$T$ & = & total thrust of the multirotor helicopter, N\\
$L,M,N$ & = & airframe roll, pitch and yaw torque of the multirotor
helicopter, N$\cdot$m\\
$m_{a}$ & = & mass of the multirotor helicopter, kg\\
$g$ & = & acceleration of gravity, kg$\cdot$m/s$^{2}$\\
$J_{x},J_{y},J_{z}$ & = & moment of inertia around the roll, pitch and yaw
axes of the\\
&  & multirotor helicopter frame, kg$\cdot$m$^{2}$\\
$f_{i}$ & = & lift of the $i$-th rotor, N\\
$K$ & = & maximum lift of each rotor, N\\
$\eta_{i}$ & = & efficiency parameter of the $i$-th rotor\\
$d$ & = & distance from the center of the rotor to the center of mass\\
$k_{\mu}$ & = & ratio between the reactive torque and the lift of the rotors
\end{tabular}

\section{Introduction}

Multirotor helicopters are attracting increasing attention in recent years
because of their important contribution and cost effective application in
several tasks such as surveillance, search and rescue missions and so on.
However, there exists a potential risk to civil safety if the helicopters
crash especially in an urban area. Therefore, it is of great importance to
consider the flight safety of multirotor helicopters in the presence of rotor
faults or failures.

Over-actuated aircraft have the potential to improve safety and reliability.
Fault tolerant control of over-actuated aircraft subject to actuator failures
is discussed widely \cite{Bib Youmin Zhang}\cite{Advances Agneta}%
\cite{Fault-tolerant Guillaume}. Most works on fault tolerant control
implicitly assume that the control systems are still controllable in the event
of failures. However, few works consider the controllability of the systems
with faults. If the system is uncontrollable, any fault tolerant control
strategy will be unavailable. In \cite{imav2012}, Schneider \emph{et al}.
proposed a useful method to study the controllability of multirotor systems
with rotor failures based on the construction of the attainable control set.
However, they did not give theoretical analysis of the controllability of the
multirotor systems. This is one of our motivations.

Sometimes, a hexacopter is uncontrollable if one rotor fails. Owing to this,
the hexacopter subject to rotor failures is often controlled by leaving the
yaw states uncontrolled, the feasibility of which has been tested by
\cite{imav2012}\cite{TEDquad}. This is very useful in emergency situations.
However, we find that not all the uncontrollable hexacopters can be controlled
by the degraded way mentioned in \cite{imav2012}\cite{TEDquad}. If the maximum
lift of each rotor is lower than a certain value then the degraded system,
where the yaw states of the considered hexacopter are ignored, is still
uncontrollable. Our another motivation is to specify this lower bound value.

In this paper, the controllability of a class of hexacopters subject to one
rotor failure is analyzed based on the positive controllability theory in
\cite{Brammer(1972)}, and the results show that the hexacopter is
uncontrollable. In order to land the hexacopter safely, a Degraded Control
Strategy (DCS) is proposed for the degraded system. The lower bound of the
maximum lift of each rotor is specified, which can help the designers in
choosing the proper rotors for improving the fault-tolerant capability of the
hexacopter. The major contributions of this paper are: (i) a theoretical
controllability analysis for a class of hexacopters, and (ii) the
specification of the lower bound of the maximum rotor lift.

\section{Problem Formulation}

\subsection{Hexacopter Model}

This paper considers a class of PNPNPN hexacopters shown in Fig.1. The linear
dynamical model around hover conditions is given as follows \cite{Du-ASDB}%
\cite{Guillaume(2011)}:%
\begin{equation}
\dot{x}=Ax+B\underset{u}{\underbrace{\left(  F-G\right)  }}
\label{hexacopter_model}%
\end{equation}
where%
\begin{align*}
x  &  =\left[  h\text{ }\phi\text{ }\theta\text{ }\psi\text{ }v_{h}\text{
}p\text{ }q\text{ }r\right]  ^{T}\in%
\mathbb{R}
^{8},F=\left[  T\text{ }L\text{ }M\text{ }N\right]  ^{T}\in%
\mathbb{R}
^{4},G=\left[  m_{a}g\text{ }0\text{ }0\text{ }0\right]  ^{T}\in%
\mathbb{R}
^{4},\\
A  &  =%
\begin{bmatrix}
0_{4\times4} & I_{4}\\
0 & 0
\end{bmatrix}
\in%
\mathbb{R}
^{8\times8},B=%
\begin{bmatrix}
0\\
J_{f}^{-1}%
\end{bmatrix}
\in%
\mathbb{R}
^{8\times4},J_{f}=\text{diag}\left(  -m_{a},J_{x},J_{y},J_{z}\right)
\end{align*}
and
\begin{equation}
u=F-G\in\mathcal{U}\subset%
\mathbb{R}
^{4}. \label{u}%
\end{equation}
\begin{figure}[ptb]
\begin{center}
\includegraphics[
scale=0.7 ]{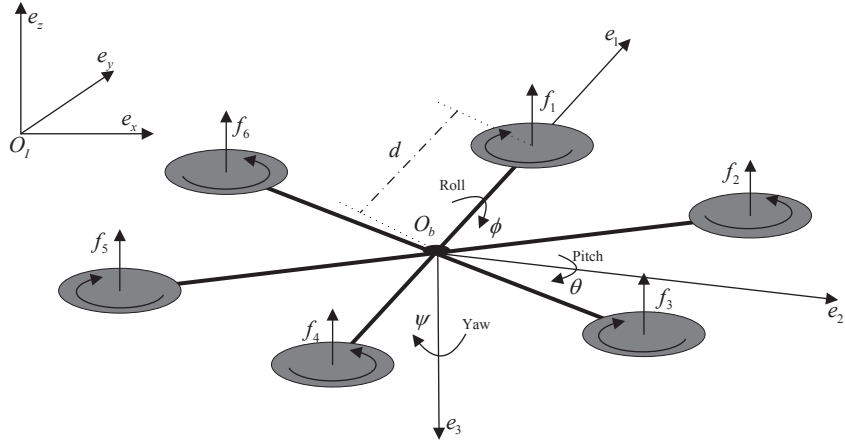}
\end{center}
\caption{Kinematic scheme of a PNPNPN hexacopter, where P denotes that a rotor
rotates clockwise and N denotes that a rotor rotates anticlockwise}%
\label{Hexacopter_configuration}%
\end{figure}

According to the geometry of the hexacopter shown in
Fig.\ref{Hexacopter_configuration}, the mapping from the rotor lift
$f_{i},i=1,\cdots,6$ to the system total thrust/torque $F$ is \cite{imav2012}%
\cite{Du-ASDB}:
\begin{equation}
F=H_{\eta_{1},\cdots,\eta_{6}}f \label{F=Hf}%
\end{equation}
where $f=\left[  f_{1}\text{ }\cdots\text{ }f_{6}\right]  ^{T}\in%
\mathbb{R}
^{6}$ and the control effectiveness matrix $H_{\eta_{1},\cdots,\eta_{6}}\in%
\mathbb{R}
^{4\times6}$ in parameterized form is%
\begin{equation}
H_{\eta_{1},\cdots,\eta_{6}}=\left[
\begin{array}
[c]{cccccc}%
\eta_{1} & \eta_{2} & \eta_{3} & \eta_{4} & \eta_{5} & \eta_{6}\\
0 & -\frac{\sqrt{3}}{2}\eta_{2}d & -\frac{\sqrt{3}}{2}\eta_{3}d & 0 &
\frac{\sqrt{3}}{2}\eta_{5}d & \frac{\sqrt{3}}{2}\eta_{6}d\\
\eta_{1}d & \frac{1}{2}\eta_{2}d & -\frac{1}{2}\eta_{3}d & -\eta_{4}d &
-\frac{1}{2}\eta_{5}d & \frac{1}{2}\eta_{6}d\\
-\eta_{1}k_{\mu} & \eta_{2}k_{\mu} & -\eta_{3}k_{\mu} & \eta_{4}k_{\mu} &
-\eta_{5}k_{\mu} & \eta_{6}k_{\mu}%
\end{array}
\right]  . \label{H}%
\end{equation}
The parameter $\eta_{i}\in\left[  0,1\right]  ,i=1,\cdots,6$ is used to
account for rotor wear/failure. If the $i$-th rotor fails, then $\eta_{i}=0$.
Since the rotors of the hexacopter can only provide upward lifts, we let
$f_{i}\in\left[  0,K\right]  ,i=1,\cdots,6$. As a result, we have
\begin{equation}
f\in\mathcal{F}=\Pi_{i=1}^{6}\left[  0,K\right]  . \label{f_constrain}%
\end{equation}

\subsection{Control Constraint}

In this section, we will specify the control constraint $\mathcal{U}.$
Combining (\ref{u}), (\ref{F=Hf}) and (\ref{f_constrain}), we can get the
control constraint%
\begin{equation}
\mathcal{U}_{\eta_{1},\cdots,\eta_{6}}^{0}=\left\{  u|u=H_{\eta_{1}%
,\cdots,\eta_{6}}f-G,f\in\mathcal{F}\right\}  . \label{U0}%
\end{equation}
Next, we consider the control constraint $\mathcal{U}$ under a control
allocation. In practice, the virtual control $F$ is often designed first.
Then, the control allocation is used to obtain $f$ as%
\begin{equation}
f=P_{\eta_{1},\cdots,\eta_{6}}F \label{f=Pu}%
\end{equation}
where $P_{\eta_{1},\cdots,\eta_{6}}\in%
\mathbb{R}
^{6\times4}$ is the allocation matrix satisfying
\begin{equation}
H_{\eta_{1},\cdots,\eta_{6}}P_{\eta_{1},\cdots,\eta_{6}}=I_{4}.
\label{allocation}%
\end{equation}
Since $F=u+G$ from (\ref{u}), we can get the control constraint $\mathcal{U}$
under the control allocation (\ref{f=Pu}) as%
\begin{equation}
\mathcal{U}_{\eta_{1},\cdots,\eta_{6}}^{a}=\left\{  u|P_{\eta_{1},\cdots
,\eta_{6}}\left(  u+G\right)  \in\mathcal{F}\right\}  . \label{Ua}%
\end{equation}
The pseudo-inverse matrix (PIM) method \cite{Guillaume(2011)}%
\cite{Oppenheimer} is often used to choose $P_{\eta_{1},\cdots,\eta_{6}}$ as
follows%
\begin{equation}
P_{\eta_{1},\cdots,\eta_{6}}=H_{\eta_{1},\cdots,\eta_{6}}^{\dagger}%
=H_{\eta_{1},\cdots,\eta_{6}}^{T}\left(  H_{\eta_{1},\cdots,\eta_{6}}%
H_{\eta_{1},\cdots,\eta_{6}}^{T}\right)  ^{-1}. \label{PIMM}%
\end{equation}

The relation between $\mathcal{U}_{\eta_{1},\cdots,\eta_{6}}^{a}$ and
$\mathcal{U}_{\eta_{1},\cdots,\eta_{6}}^{0}$ is stated as \emph{Theorem 1},
which is consistent with the results in \cite{Guillaume(2011)} and
\cite{Demenkov}.

\textbf{Theorem 1}. $\mathcal{U}_{\eta_{1},\cdots,\eta_{6}}^{a}\subseteq
\mathcal{U}_{\eta_{1},\cdots,\eta_{6}}^{0}.$

\textit{Proof}. For any $u^{\ast}\in\mathcal{U}_{\eta_{1},\cdots,\eta_{6}}%
^{a},$ there exists a $f^{\ast}\in\mathcal{F}$ such that$\ f^{\ast}%
=P_{\eta_{1},\cdots,\eta_{6}}\left(  u^{\ast}+G\right)  .$ By
(\ref{allocation}), we have $H_{\eta_{1},\cdots,\eta_{6}}f^{\ast}%
-G=H_{\eta_{1},\cdots,\eta_{6}}P_{\eta_{1},\cdots,\eta_{6}}\left(  u^{\ast
}+G\right)  -G=u^{\ast}.$ This implies $u^{\ast}\in\mathcal{U}_{\eta
_{1},\cdots,\eta_{6}}^{0},$ namely $\mathcal{U}_{\eta_{1},\cdots,\eta_{6}}%
^{a}\subseteq\mathcal{U}_{\eta_{1},\cdots,\eta_{6}}^{0}.$ $\square$

\subsection{Objective}

The first objective is to show that the system (\ref{hexacopter_model}) will
lose controllability\footnote{The system (\ref{hexacopter_model}) with
constraint set $\mathcal{U}\subset%
\mathbb{R}
^{4}$ is called controllable if, for each pair of points $x_{0}\in%
\mathbb{R}
^{8}$ and $x_{1}\in%
\mathbb{R}
^{8}$, there exists a bounded admissible control, $u\left(  t\right)
\in\mathcal{U}$, defined on some finite interval $0\leq t\leq t_{1}$, which
steers $x_{0}$ to $x_{1}$. Specifically, the solution to
(\ref{hexacopter_model}), $x\left(  t,u\left(  \cdot\right)  \right)  $,
satisfies the boundary conditions $x\left(  0,u\left(  \cdot\right)  \right)
=x_{0}$ and $x\left(  t_{1},u\left(  \cdot\right)  \right)  =x_{1}$.} when one
rotor fails. That is, the system (\ref{hexacopter_model}) is uncontrollable
subject to the control constraint $\mathcal{U=U}_{\eta_{i}=0}^{0}$ where, for
simplicity, the subscript $\eta_{i}=0$ is used to denote that only the $i$-th
rotor fails and the remaining rotors have neither wear nor failures. The
second objective is to study the controllability of the degraded system, where
the yaw states are removed from (\ref{hexacopter_model}), and specify the
lower bound of the maximum lift of each rotor.

\textbf{Remark 1.} Not all the hexacopters are configured as Fig.1. For
example, a class of PPNNPN hexacopters are considered in \cite{imav2012}. It
is pointed out that other type of hexacopters can be analyzed in the same way
as the popular PNPNPN hexacopter.

\textbf{Remark 2}. Classical controllability theories of linear systems often
require the origin to be an interior point of $\mathcal{U}$\ so that
$\mathcal{C}\left(  A,B\right)  \ $being row full rank is a necessary and
sufficient condition \cite{Brammer(1972)}. However, for the system
(\ref{hexacopter_model}) the control constraint $\mathcal{U=U}_{\eta_{i}%
=0}^{0}$ does not have the origin as its interior point when some rotors are
damaged or fail. Consequently, $\mathcal{C}\left(  A,B\right)  $ being row
full rank is not sufficient to test the controllability of the system
(\ref{hexacopter_model}).

\section{Controllability of the Hexacopter Subject to One Rotor Failure}

\subsection{Preliminaries}

In this section, we will study the controllability of the hexacopter subject
to one rotor failure based on the positive controllability theory proposed in
\cite{Brammer(1972)}. Applying the positive controllability theorem in
\cite{Brammer(1972)} to the system (\ref{hexacopter_model}) directly, we have

\textbf{Theorem 2}. Consider the system (\ref{hexacopter_model}), suppose that
the set $\mathcal{U}$ contains a vector in the kernel of $B$ (i.e., there
exists $u\in\mathcal{U}$ satisfying $Bu=0$) and the set $\mathcal{CH}\left(
\mathcal{U}\right)  $\footnote{$\mathcal{CH}\left(  \mathcal{U}\right)  $ is
the convex hull of $\mathcal{U}$. According to \cite{Boyd 2004}, the convex
hull of $\Delta$ is the set of all convex combinations of points in $\Delta$.
If $\Delta$ is convex, then $\mathcal{CH}\left(  \Delta\right)  =\Delta$.} has
nonempty interior in $%
\mathbb{R}
^{4}$. Then, the following conditions are necessary and sufficient for the
controllability of (\ref{hexacopter_model}):

\begin{enumerate}
\item[(c1)] Rank $\mathcal{C}\left(  A,B\right)  =8$, where $\mathcal{C}%
\left(  A,B\right)  =\left[  B\text{ }AB\text{ }\cdots\text{ }A^{7}B\right]
.$

\item[(c2)] There is no non-zero real eigenvector $v$ of $A^{T}$ satisfying
$v^{T}Bu\leq0$ for all $u\in\mathcal{U}.$
\end{enumerate}

For the considered linear hexacopter model (\ref{hexacopter_model}),
\emph{Theorem 2 }is simplified as follows.

\textbf{Corollary 1}. The system (\ref{hexacopter_model}) is controllable if
and only if%
\begin{equation}
\underset{v\in\mathcal{V}}{\min}\text{ }\underset{u\in\mathcal{U}}{\max}\text{
}v^{T}Bu>0 \label{min-max}%
\end{equation}
where $\mathcal{V}=\left\{  v|A^{T}v=0,\left\Vert v\right\Vert =1,v\in%
\mathbb{R}
^{8}\right\}  $.

\emph{Proof:} The proof is straightforward. For the system
(\ref{hexacopter_model}), it is easy to check that rank $\mathcal{C}\left(
A,B\right)  =8$. According to \emph{Theorem 2}, then the system
(\ref{hexacopter_model}) is controllable if and only if there is no non-zero
real eigenvector $v$ of $A^{T}$ satisfying $v^{T}Bu\leq0$ for all
$u\in\mathcal{U}$. Since all the eigenvalues of $A^{T}$ are zero, all the real
eigenvectors of $A^{T}$ can be obtained by solving linear equations $A^{T}%
v=0$. Then the system (\ref{hexacopter_model}) is controllable if and only if
(\ref{min-max}) is satisfied. The constraint $\left\Vert v\right\Vert =1$ is
used to make (\ref{min-max}) verifiable, which does not change the sign of
$v^{T}Bu$. $\square$

\subsection{Controllability Analysis of the Hexacopter Subject to One Rotor
Failure}

For the controllability of the hexacopter subject to one rotor failure, we
have the following theorem:

\textbf{Theorem 3.} The system (\ref{hexacopter_model}) constrained by
$\mathcal{U=U}_{\eta_{i}=0}^{0},\forall i\in\left\{  1,2,3,4,5,6\right\}  $ is uncontrollable.

\emph{Proof:} This proof is accomplished by counterexamples. For each
$\eta_{i}=0$, we find a vector $\hat{v}_{i}\in V$ satisfying%
\begin{equation}
\underset{u\in\mathcal{U}_{\eta_{i}=0}^{0}}{\max\text{ }}\hat{v}_{i}^{T}Bu=0
\label{counterexample}%
\end{equation}
Then%
\[
\underset{v\in\mathcal{V}}{\min}\text{ }\underset{u\in\mathcal{U}}{\max}\text{
}v^{T}Bu\leq0.
\]
Consequently, the system (\ref{hexacopter_model}) constrained by
$\mathcal{U=U}_{\eta_{i}=0}^{0},\forall i\in\left\{  1,2,3,4,5,6\right\}  $ is
uncontrollable according to to \emph{Corollary 1}. See Appendix A for details.
$\square$

As analyzed above, the PNPNPN hexacopter subject to one rotor failure is
uncontrollable. A question follows consequentially: how a hexacopter can land
safely after one rotor fails. In \cite{imav2012}\cite{TEDquad}, the author
suggested a degraded control way that was to leave the yaw states
uncontrolled. However, neither a controllability analysis nor a concrete DCS exists.

\section{Degraded Control and Analysis for Safe Landing Without Yaw}

According to Section III, the yaw states of the hexacopter may be left
uncontrolled for safe landing when one rotor fails. In this section, a DCS for
the case with one of $\eta_{i}$, $i=1,\cdots,6$ being zero is approached,
which does not require any change on the original controller. Furthermore, it
is shown that the hexacopter subject to one rotor failure can land by the DCS
if and only if the maximum lift of each rotor is greater than a certain value.
This lower bound value will be specified in this section.

\subsection{DCS for Safe Landing Without Yaw Control}

In practice, the virtual control $F$ is often designed first. Then if no rotor
fails, $f$ is obtained by $f=PF$ where $F=\left[  T\text{ }L\text{ }M\text{
}N\right]  ^{T}$ and $P$ is expressed by (\ref{PIMM}). If one of $\eta_{i}$,
$i=1,\cdots,6$ is zero, the DCS for the system (\ref{hexacopter_model})
includes the following two steps:

\emph{Step 1}\textit{:} Leave the yaw states uncontrolled. One simple way is
to let $\left(  \psi_{s},r_{s}\right)  =\left(  \psi_{c},r_{c}\right)  $,
where $\left(  \psi_{s},r_{s}\right)  $ are the sensed yaw states and $\left(
\psi_{c},r_{c}\right)  $ the commanded yaw states.

\emph{Step 2}\textit{:} Reallocate $\bar{F}$ to the set of rotor lifts $f$ by%
\begin{align}
f  &  =\bar{P}\bar{F},\label{f=P-u}\\
\bar{P}_{\eta_{1},\cdots,\eta_{6}}  &  =\bar{H}_{\eta_{1},\cdots,\eta_{6}}%
^{T}\left(  \bar{H}_{\eta_{1},\cdots,\eta_{6}}\bar{H}_{\eta_{1},\cdots
,\eta_{6}}^{T}\right)  ^{-1} \label{P-}%
\end{align}
where $\bar{F}=\left[  T\text{ }L\text{ }M\right]  ^{T}$ and%

\begin{equation}
\bar{H}_{\eta_{1},\cdots,\eta_{6}}=\left[
\begin{array}
[c]{cccccc}%
\eta_{1} & \eta_{2} & \eta_{3} & \eta_{4} & \eta_{5} & \eta_{6}\\
0 & -\frac{\sqrt{3}}{2}\eta_{2}d & -\frac{\sqrt{3}}{2}\eta_{3}d & 0 &
\frac{\sqrt{3}}{2}\eta_{5}d & \frac{\sqrt{3}}{2}\eta_{6}d\\
\eta_{1}d & \frac{1}{2}\eta_{2}d & -\frac{1}{2}\eta_{3}d & -\eta_{4}d &
-\frac{1}{2}\eta_{5}d & \frac{1}{2}\eta_{6}d
\end{array}
\right]  . \label{H--}%
\end{equation}

However, there is no theoretical analysis of the DCS in the existing
literatures according to our knowledge. In the following section, the lower
bound of the maximum lift of each rotor is specified through controllability analysis.

\subsection{Controllability Analysis of the Hexacopter Removing the Yaw
States}

The degraded system that the yaw states are removed from
(\ref{hexacopter_model}) is given as%
\begin{equation}
\dot{x}^{\ast}=\bar{A}x^{\ast}+\bar{B}\underset{\bar{u}}{\underbrace{\left(
\bar{F}-\bar{G}\right)  }} \label{linear_degraded}%
\end{equation}
where%
\begin{align*}
x^{\ast}  &  =\left[  h\text{ }\phi\text{ }\theta\text{ }v_{h}\text{ }p\text{
}q\right]  ^{T}\in%
\mathbb{R}
^{6},\bar{F}=\left[  T\text{ }L\text{ }M\right]  ^{T}\in%
\mathbb{R}
^{3},\bar{G}=\left[  m_{a}g\text{ 0 0}\right]  ^{T}\in%
\mathbb{R}
^{3},\\
\bar{A}  &  =\left[
\begin{array}
[c]{cc}%
0_{3\times3} & I_{3}\\
0 & 0
\end{array}
\right]  \in%
\mathbb{R}
^{6\times6},\bar{B}=\left[
\begin{array}
[c]{c}%
0\\
\bar{J}_{f}^{-1}%
\end{array}
\right]  \in%
\mathbb{R}
^{6\times3},\bar{J}_{f}=\text{diag}\left(  -m_{a},J_{x},J_{y}\right)
\end{align*}
and%
\[
\bar{u}=\bar{F}-\bar{G}\in\mathcal{\bar{U}}\subset%
\mathbb{R}
^{3}.
\]

Similar to the system (\ref{hexacopter_model}), the control constraint
$\mathcal{\bar{U}}$ is%
\begin{equation}
\mathcal{\bar{U}}_{\eta_{1},\cdots,\eta_{6}}^{0}=\left\{  \bar{u}|\bar{u}%
=\bar{H}_{\eta_{1},\cdots,\eta_{6}}f-\bar{G},f\in\mathcal{F}\right\}  .
\label{U0bar}%
\end{equation}
and the control constraint $\mathcal{\bar{U}}$ under the control allocation
(\ref{P-}) is%
\begin{equation}
\mathcal{\bar{U}}_{\eta_{1},\cdots,\eta_{6}}^{a}=\left\{  \bar{u}|\bar
{P}_{\eta_{1},\cdots,\eta_{6}}\left(  \bar{u}+\bar{G}\right)  \in
\mathcal{F}\right\}  . \label{Uabar}%
\end{equation}
Similar to \emph{Theorem 1}, we have $\mathcal{\bar{U}}_{\eta_{1},\cdots
,\eta_{6}}^{a}\subseteq\mathcal{\bar{U}}_{\eta_{1},\cdots,\eta_{6}}^{0}$.

Similarly to \emph{Corollary 1}, the following theorem is obtained:

\textbf{Theorem 4.} The system (\ref{linear_degraded}) constrained by
$\mathcal{\bar{U}}$ is controllable if and only if%
\begin{equation}
\underset{v\in\mathcal{\bar{V}}}{\min}\underset{\bar{u}\in\mathcal{\bar{U}}%
}{\text{ }\max\text{ }}\bar{v}^{T}\bar{B}\bar{u}>0 \label{min-max-degraded}%
\end{equation}
where $\mathcal{\bar{V}}=\left\{  v|\bar{A}^{T}v=0,\left\Vert v\right\Vert
=1,v\in%
\mathbb{R}
^{6}\right\}  $.

\emph{Proof:} This proof is similar to the proof of \emph{Corollary 1}. See
Appendix B for details. $\square$

\textbf{Theorem 5}. The system (\ref{linear_degraded}) constrained by
$\mathcal{\bar{U}=\bar{U}}_{\eta_{i}=0}^{a},\forall i\in\left\{
1,2,3,4,5,6\right\}  $ is controllable if and only if%
\begin{equation}
K>\frac{5}{18}m_{a}g. \label{K}%
\end{equation}
Furthermore, the system (\ref{linear_degraded}) constrained by $\mathcal{\bar
{U}=\bar{U}}_{\eta_{i}=0}^{0},\forall i\in\left\{  1,2,3,4,5,6\right\}  $ is
controllable if (\ref{K}) holds.

\emph{Proof:} Under $\mathcal{\bar{U}=\bar{U}}_{\eta_{2}=0}^{a}$ we first
prove that the following two propositions hold (see Appendix C).

\emph{Proposition 1:} there is a $\bar{v}_{2}\in\mathcal{\bar{V}}$ satisfying%
\begin{equation}
\underset{u\in\mathcal{U}_{\eta_{2}=0}^{0}}{\max\text{ }}\bar{v}_{2}^{T}%
\bar{B}\bar{u}\leq0 \label{max_v2}%
\end{equation}
if $K\leq\frac{5}{18}m_{a}g$.

\emph{Proposition 2:} there is no such a $\bar{v}_{2}\in\mathcal{\bar{V}}$
satisfying (\ref{max_v2}) if $K>\frac{5}{18}m_{a}g$.

With \emph{Proposition 1}\ and \emph{Proposition 2}, the system
(\ref{linear_degraded}) constrained by $\mathcal{\bar{U}=\bar{U}}_{\eta_{2}%
=0}^{a}$ is controllable if and only if (\ref{K}) holds according to
\emph{Theorem 4}. If (\ref{K}) holds, then for each pair of points $\bar
{x}_{0}\in%
\mathbb{R}
^{6}$ and $\bar{x}_{1}\in%
\mathbb{R}
^{6}$ there exists a $\bar{u}^{\ast}\left(  t\right)  \in\mathcal{\bar{U}%
}_{\eta_{2}=0}^{a}$, which steers $x_{0}$ to $x_{1}$. Since $\mathcal{\bar{U}%
}_{\eta_{2}=0}^{a}\subseteq\mathcal{\bar{U}}_{\eta_{2}=0}^{0}$, $\bar{u}%
^{\ast}\left(  t\right)  \in\mathcal{\bar{U}}_{\eta_{2}=0}^{0}$, namely the
system (\ref{linear_degraded}) constrained by $\mathcal{\bar{U}=\bar{U}}%
_{\eta_{2}=0}^{0}$ is controllable. Similarly, we can conclude this proof for
$\mathcal{\bar{U}=\bar{U}}_{\eta_{i}=0}^{a},\forall i\in\left\{
1,3,4,5,6\right\}  $. $\square$

\textbf{Remark 3. }According to \emph{Theorem 5}, the designers should choose
proper rotors satisfying $K>\frac{5}{18}m_{a}g$ so as to make sure that the
hexacopter can adopt the DCS proposed in this paper.

\section{Simulation and Experiment}

\begin{figure}[ptb]
\begin{center}
\includegraphics[
scale=0.3 ]{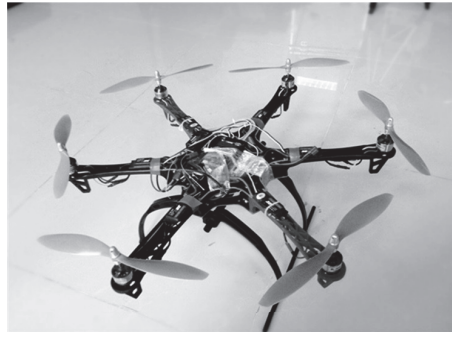}
\end{center}
\caption{A prototype hexacopter}%
\label{Hexacopter_experiment}%
\end{figure}

In order to show the feasibility of the proposed DCS, simulations and an
experiment of a prototype hexacopter (see Fig.\ref{Hexacopter_experiment}) are
carried out. The physical parameters of the prototype hexacopter are shown in
Table \ref{parameters}. In the simulation, the hexacopter is controlled by
Proportional-Derivative (PD) controllers and the proposed DCS for safe landing
is applied. After $\eta_{2}=0$, the hexacopter keeps its $\left(
h,\phi,\theta\right)  $ to the desired targets by leaving the yaw states
uncontrolled. In the experiment, a real flight test for the prototype
hexacopter was carried out. During the real flight test, $\eta_{2}$ was set to
zero. Then the DCS for safe landing kept the hexacopter level and the
hexacopter was landed by the remote-controller avoiding loss of
control.

\begin{table}[tbh]
\caption{HEXACOPTER PARAMETERS}%
\label{parameters}%
\centering%
\begin{tabular}
[c]{|l|l|l|l|}\hline
Parameter & Description & Value & Units\\\hline
m & Mass & 1.535 & kg\\\hline
g & Gravity & 9.80 & m/$s^{2}$\\\hline
d & Rotor to mass center distance & 0.275 & m\\\hline
K & Maximum lift of each rotor & 6.125 & N\\\hline
$J_{x}$ & Moment of inertia & 0.0411 & $kg.m^{2}$\\\hline
$J_{y}$ & Moment of inertia & 0.0478 & $kg.m^{2}$\\\hline
$J_{z}$ & Moment of inertia & 0.0599 & $kg.m^{2}$\\\hline
$k_{\mu}$ & $k/\mu$ & 0.1 & -\\\hline
\end{tabular}
\end{table}

\subsection{Simulation Results}

Based on the parameters in Table \ref{parameters}, a digital simulation is
performed. The hexacopter hovers at $h_{c}=1$ m and $\left[  \phi_{c}\text{
}\theta_{c}\text{ }\psi_{c}\right]  ^{T}=\left[  0\text{ }0\text{ }5\right]
^{T}$ rad controlled by Proportional-Derivative (PD) controllers which are
expressed by%
\begin{align}
L  &  =20\left(  \phi-\phi_{c}\right)  +3p,M=20\left(  \theta-\theta
_{c}\right)  +3q,\nonumber\\
N  &  =20\left(  \psi-\psi_{c}\right)  +3r,T=10\left(  h-h_{c}\right)
+6v_{h}+m_{a}g. \label{PD}%
\end{align}
If no rotor fails, $f$ is obtained by%
\begin{equation}
f=H_{\eta_{1},\cdots,\eta_{6}}^{T}\left(  H_{\eta_{1},\cdots,\eta_{6}}%
H_{\eta_{1},\cdots,\eta_{6}}^{T}\right)  ^{-1}F \label{PIM}%
\end{equation}
where $F=\left[  T\text{ }L\text{ }M\text{ }N\right]  ^{T}.$ And if one of
$\eta_{i}$, $i\in\left\{  1,2,3,4,5,6\right\}  $ is zero, $f$ is obtained by%
\begin{equation}
f=\bar{H}_{\eta_{i}=0}^{T}\left(  \bar{H}_{\eta_{i}=0}\bar{H}_{\eta_{i}=0}%
^{T}\right)  ^{-1}\bar{F} \label{PIM_degraded}%
\end{equation}
where $\bar{F}=\left[  T\text{ }L\text{ }M\right]  ^{T}.$

\begin{figure}[ptb]
\begin{center}
\includegraphics[
scale=0.7 ]{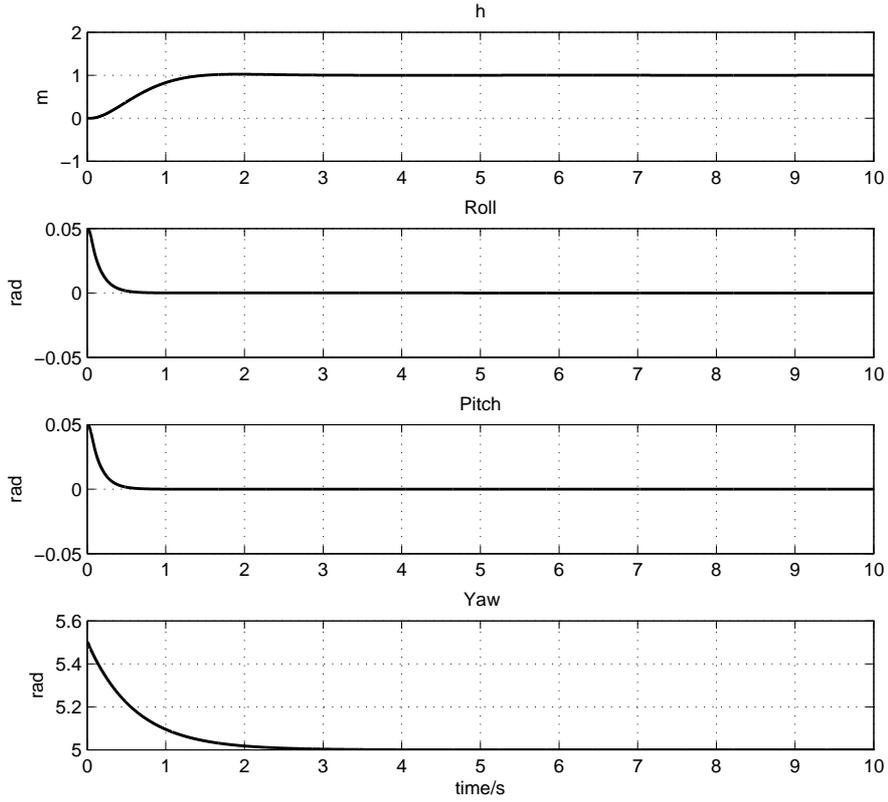}
\end{center}
\caption{No rotor fails and $h,\phi,\theta,\psi$ are controlled to the desired
target}%
\label{Figure_a}%
\end{figure}

\begin{figure}[ptb]
\begin{center}
\includegraphics[
scale=0.7 ]{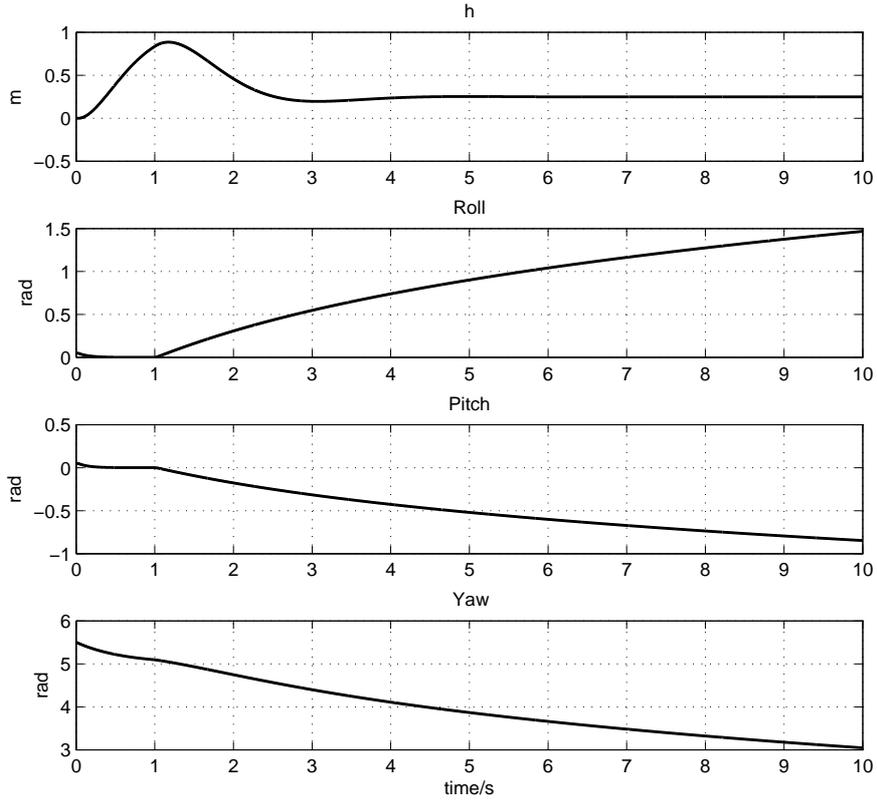}
\end{center}
\caption{The DCS is not adopted after\ $\eta_{2}=0$ and $h,\phi,\theta,\psi$
diverge from their target}%
\label{Figure_b}%
\end{figure}

\begin{figure}[ptb]
\begin{center}
\includegraphics[
scale=0.7 ]{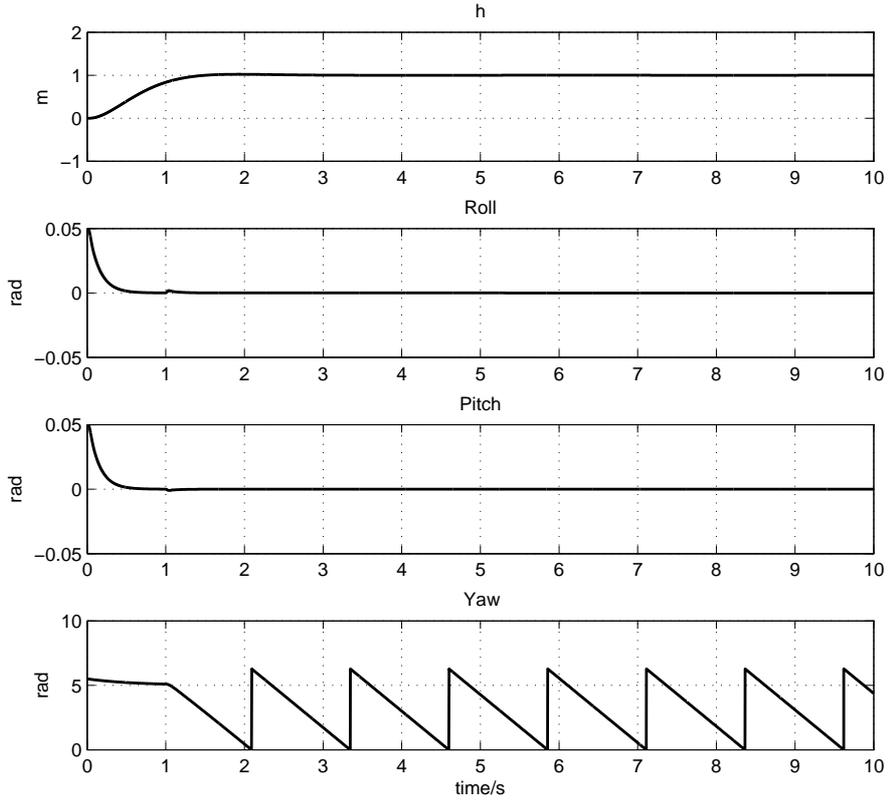}
\end{center}
\caption{The DCS is adopted after\ $\eta_{2}=0$ and $h,\phi,\theta,\psi$ are
controlled to the desired target with nice performance}%
\label{Figure_c}%
\end{figure}

Fig.\ref{Figure_a} shows the simulation results when no rotor fails, where
$h,\phi,\theta$, and $\psi$ are controlled to the desired target with nice
performance. At time instant $t=1s$, $\eta_{2}$ is set to $0$.
Fig.\ref{Figure_b} shows the simulation results when $\eta_{2}=0$ and the DCS
is not adopted. It is shown that $h,\phi,\theta$, and $\psi$ diverge from
their targets. With the DCS, $h,\phi$, and $\theta$ are controlled to the
desired targets with nice performance (see Fig.\ref{Figure_c}) which avoids
loss of control.

\begin{figure}[ptb]
\begin{center}
\includegraphics[
scale=0.7 ]{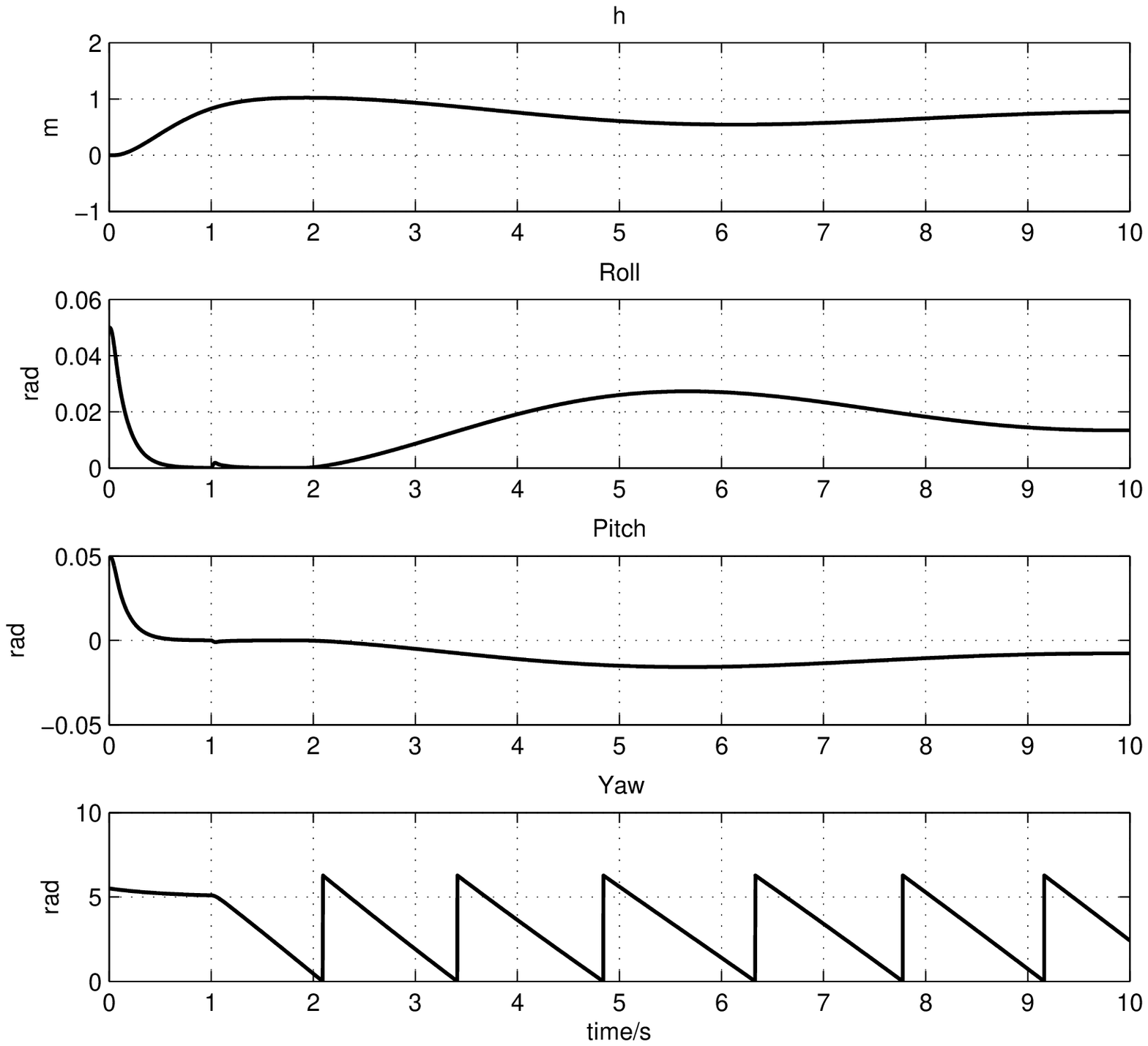}
\end{center}
\caption{$K=\frac{4.9}{18}m_{a}g$ and the hexacopter is out of control even
though the DCS is adopted}%
\label{Figure_d}%
\end{figure}

According to \emph{Theorem 5}, not all the uncontrollable hexacopters can land
in the degraded way proposed in this paper. It should be pointed out that if
$K\leq\frac{5}{18}m_{a}g$, then $h,\phi$, and $\theta$ are not controllable
and the hexacopter will crash to the land if one rotor fails. In the
simulation, we change the value of $K$ to $\frac{4.9}{18}m_{a}g$ and the
simulation results of $h,\phi,\theta$ are shown in Fig.\ref{Figure_d} where
the DCS is adopted. Obviously, the hexacopter is out of control.

\textbf{Remark 4}. In Fig.\ref{Figure_c}, the yaw angle changes with a
constant angular velocity at last. When the hexacopter rotates fast, the
damping moment $N_{D}=K_{ND}r^{2}$ can not be ignored, where $K_{ND}$ is the
damping coefficient. In the simulation we choose $K_{ND}=0.2$N$\cdot
$m/rad$^{2}$ to make the simulation results be consistent with the experiment
results. Parameters $\eta_{i}$, $i=1,\cdots,6$, which in practice can be
obtained by fault diagnosis strategies \cite{YMZ-FDD}\cite{YMZ-TSKF}, are
assumed to be known. Since the effect of fault diagnosis strategies are not in
the scope of this paper, they will not be discussed here and will be
invertigated in our future researches.

\subsection{Experimental Results}

\begin{figure}[ptb]
\begin{center}
\includegraphics[
scale=0.7 ]{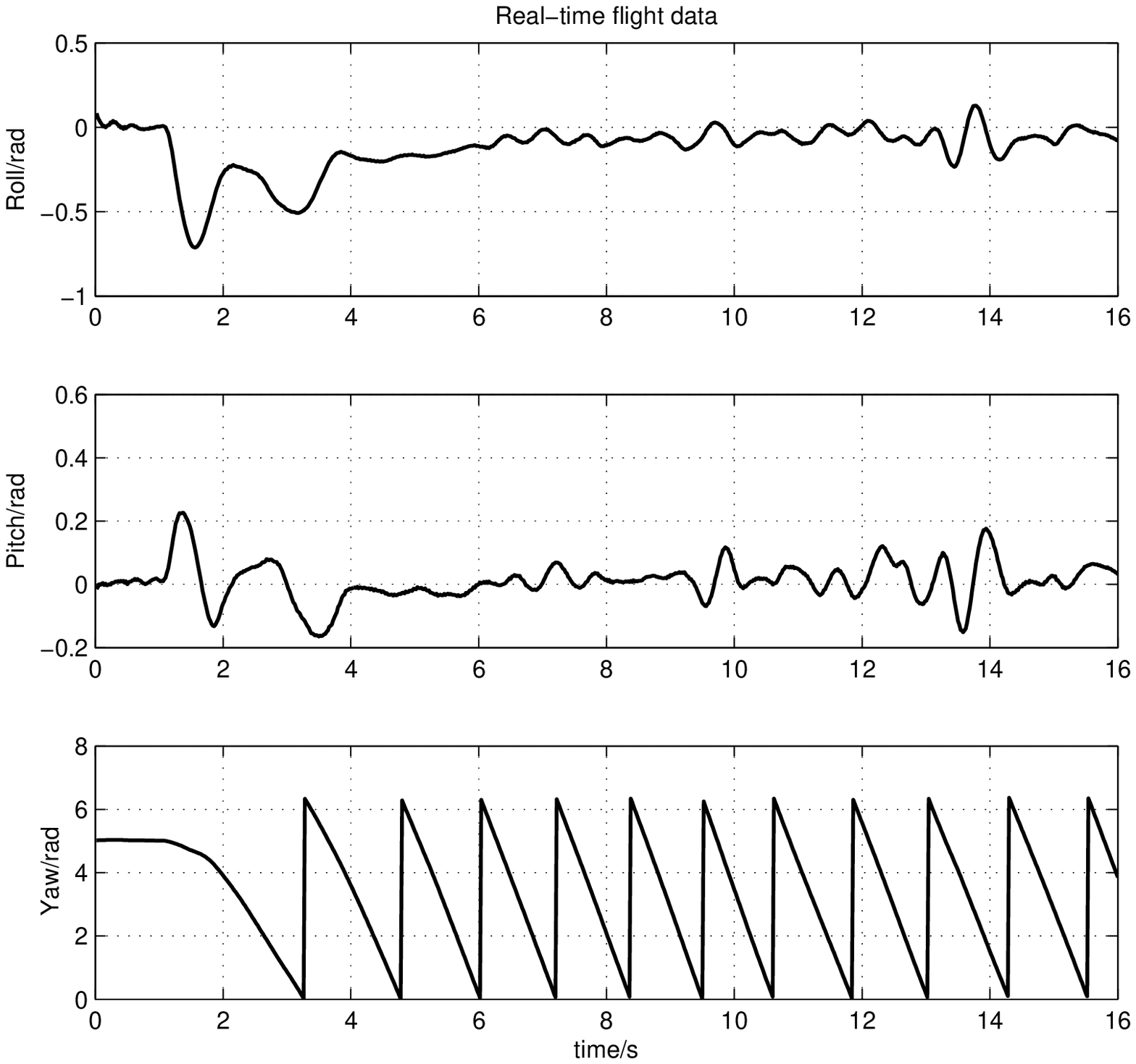}
\end{center}
\caption{Real-time flight test for a prototype hexacopter}%
\label{LAND}%
\end{figure}

In order to show the feasibility of the proposed DCS, a real flight test of
the prototype hexacopter shown in Fig.\ref{Hexacopter_experiment} was carried
out. During the flight, $\left[  \phi_{c}\text{ }\theta_{c}\text{ }\psi
_{c}\right]  ^{T}=\left[  0\text{ }0\text{ }5\right]  ^{T}$ rad and $h$ was
controlled by a remote-controller. Part of the flight data is shown in
Fig.\ref{LAND}. The hexacopter was in a stabilize mode (where $\phi
,\theta,\psi$ were controlled by Proportional-Integral-Derivative controllers
and $h$ was controlled by a remote-controller) before time $t=1s$. At time
instant $t=1s$, $\eta_{2}$ was set to $0$, then the controller kept
$\phi,\theta$ around zero by the DCS. And the hexacopter was landed slowly by
the remote-controller avoiding a flight crash.

\textbf{Remark 5}. According to Fig.\ref{LAND}, the hexacopter rotates fast
(nearly $2\pi$ rad/s) after $\eta_{2}=0$. But the $h$ can be controlled by the
remote-controller to achieve a safe landing. The video of the experiment is
online \cite{youtube}.

\section{Conclusions}

In this paper, the controllability and fault tolerant control problem of a
class of hexacopters are investigated. The following two conclusions are
obtained: i) although the considered hexacopter is over-actuated and its
controllability matrix is row full rank, it is uncontrollable when one rotor
fails, and ii) the uncontrollable hexacopter can land in a degraded way by the
proposed Degraded Control Strategy (DCS) under the condition that the maximum
lift of each rotor is greater than $\frac{5}{18}$ of the hexacopter's gravity.
The simulation and experiment results on a prototype hexacopter show the
feasibility of the proposed DCS. The focus of our future work is to extend the
controllability theory in this paper to analyze the controllability of general
multirotor helicopters.

\section{Appendix}

\subsection{Proof of Theorem 3}

This proof is accomplished by counterexamples.

(i) Case $\eta_{2}=0$. The control effectiveness matrix $H_{\eta_{2}=0}$ is
expressed by%
\begin{equation}
H_{\eta_{2}=0}=\left[
\begin{array}
[c]{cccccc}%
1 & 0 & 1 & 1 & 1 & 1\\
0 & 0 & -\frac{\sqrt{3}}{2}d & 0 & \frac{\sqrt{3}}{2}d & \frac{\sqrt{3}}{2}d\\
d & 0 & -\frac{1}{2}d & -d & -\frac{1}{2}d & \frac{1}{2}d\\
-k_{\mu} & 0 & -k_{\mu} & k_{\mu} & -k_{\mu} & k_{\mu}%
\end{array}
\right]  . \label{H2}%
\end{equation}
By solving $H_{\eta_{2}=0}f=F$ based on the theory of linear algebra
\cite{Greub(1967)}, $\mathcal{U}_{\eta_{2}=0}^{0}=\left\{  u|u=\left[
T-m_{a}g\text{ }L\text{ }M\text{ }N\right]  ^{T}\right\}  $ is given by the
following inequalities
\begin{subequations}
\label{U2}%
\begin{align}
0  &  \leq\frac{1}{2}T+\frac{2}{3d}M+\frac{1}{6k_{\mu}}N-\alpha\leq
K\label{1}\\
0  &  \leq-\frac{\sqrt{3}}{3d}L-\frac{1}{3d}M-\frac{1}{3k_{\mu}}N+\alpha\leq
K\label{2}\\
0  &  \leq\frac{1}{2}T+\frac{1}{2d}N-\alpha\leq K\label{3}\\
0  &  \leq\frac{\sqrt{3}}{3d}L-\frac{1}{3d}M-\frac{1}{3k_{\mu}}N\leq
K\label{4}\\
0  &  \leq\alpha\leq K. \label{5}%
\end{align}

Let $v_{2}=\left[  0\text{ }0\text{ }0\text{ }0\text{ }0\text{ }-\frac
{\sqrt{3}J_{x}}{3d}\text{ }\frac{J_{y}}{3d}\text{ }\frac{J_{z}}{3k_{\mu}%
}\right]  ^{T}$ and $\hat{v}_{2}=\frac{v_{2}}{\left\Vert v_{2}\right\Vert }$,
we have $\hat{v}_{2}\in\mathcal{V}$ and%
\end{subequations}
\[
\hat{v}_{2}^{T}Bu=\frac{-\frac{\sqrt{3}}{3d}L+\frac{1}{3d}M+\frac{1}{3k_{\mu}%
}N}{\left\Vert v_{2}\right\Vert }.
\]
According to (\ref{4}),
\[
\underset{u\in\mathcal{U}_{\eta_{2}=0}^{0}}{\max\text{ }}\hat{v}_{2}^{T}Bu=0.
\]

(ii) Case $\eta_{i}=0$. Similar to the case $\eta_{2}=0$, we can find a
$\hat{v}_{i}\in\mathcal{V}$ satisfying%
\[
\underset{u\in\mathcal{U}_{\eta_{i}=0}^{0}}{\max\text{ }}\hat{v}_{i}%
^{T}Bu=0,i\in\left\{  1,3,4,5,6\right\}  .
\]

From (i) and (ii), we have%
\[
\underset{v\in\mathcal{V}}{\min}\text{ }\underset{u\in\mathcal{U}}{\max}\text{
}v^{T}Bu\leq0
\]
and the system (\ref{hexacopter_model}) constrained by $\mathcal{U=U}%
_{\eta_{i}=0}^{0},\forall i\in\left\{  1,2,3,4,5,6\right\}  $ is
uncontrollable according to \emph{Corollary 1}.

\subsection{Proof of Theorem 4}

We apply the positive controllability theorem in \cite{Brammer(1972)} to the
system (\ref{linear_degraded}) directly. Suppose that the set $\mathcal{\bar
{U}}$ contains a vector in the kernel of $\bar{B}$ and the set $\mathcal{CH}%
\left(  \mathcal{\bar{U}}\right)  $ has nonempty interior in $%
\mathbb{R}
^{3}$, the following conditions are necessary and sufficient for the
controllability of (\ref{linear_degraded}):

\begin{enumerate}
\item[(i)] Rank $\mathcal{C}\left(  \bar{A},\bar{B}\right)  =6$, where
$\mathcal{C}\left(  \bar{A},\bar{B}\right)  =\left[  \bar{B}\text{ }\bar
{A}\bar{B}\text{ }\cdots\text{ }\bar{A}^{5}\bar{B}\right]  .$

\item[(ii)] There is no non-zero real eigenvector $v$ of $\bar{A}^{T}$
satisfying $v^{T}\bar{B}\bar{u}\leq0$ for all $\bar{u}\in\mathcal{\bar{U}}.$
\end{enumerate}

For the system (\ref{linear_degraded}), it is easy to check that rank
$\mathcal{C}\left(  \bar{A},\bar{B}\right)  =6$. Since all the eigenvalues of
$\bar{A}^{T}$ are zero, all the real eigenvectors of $\bar{A}^{T}$ can be
obtained by solving linear equations $\bar{A}^{T}v=0$. Then the system
(\ref{linear_degraded}) is controllable if and only if (\ref{min-max-degraded}%
) is true. $\square$

\subsection{Proof of Theorem 5}

\subsubsection{Proof of Proposition 1}

According to (\ref{P-}) and (\ref{H--}), $\bar{P}_{\eta_{2}=0}=\bar{H}%
_{\eta_{2}=0}^{T}\left(  \bar{H}_{\eta_{2}=0}\bar{H}_{\eta_{2}=0}^{T}\right)
^{-1}$. Then $\mathcal{\bar{U}}_{\eta_{2}=0}^{a}=\left\{  \bar{u}|\bar
{u}=\left[  T-m_{a}g\text{ }L\text{ }M\right]  ^{T}\right\}  $ is given by the
following inequalities according to (\ref{Uabar})%

\begin{align}
-\frac{5}{18}T  &  \leq-\frac{\sqrt{3}}{9d}L+\frac{4}{9d}M\leq K-\frac{5}%
{18}T,\nonumber\\
-\frac{5}{18}T  &  \leq-\frac{5\sqrt{3}}{18d}L-\frac{1}{18d}M\leq K-\frac
{5}{18}T,\nonumber\\
-\frac{1}{6}T  &  \leq-\frac{1}{3d}M\leq K-\frac{1}{6}T,\nonumber\\
-\frac{1}{9}T  &  \leq\frac{2\sqrt{3}}{9d}L-\frac{2}{9d}M\leq K-\frac{1}%
{9}T,\nonumber\\
-\frac{1}{6}T  &  \leq\frac{\sqrt{3}}{6d}L+\frac{1}{6d}M\leq K-\frac{1}{6}T.
\label{inequalities}%
\end{align}

Denote $E_{c}=\left\{  c|c=\left(  L,M\right)  ^{T},L,M\text{ satisfy
(\ref{inequalities})}\right\}  $ which is closed and convex. If $T\geq
\frac{18}{5}K$, $c_{0}=\left[  0\text{ }0\right]  ^{T}$ is not\ an interior
point of $E_{c}$. Then there is a non-zero vector $c_{k}=\left[  k_{c1}\text{
}k_{c2}\right]  ^{T}$ satisfying%
\begin{equation}
c_{k}^{T}\left(  c-c_{0}\right)  =k_{c1}L+k_{c2}M\leq0 \label{case2-}%
\end{equation}
for all $c=\left(  L,M\right)  ^{T}\in E_{c}$ according to \cite{Goodwin}. Let
$v_{2}=\left[  \text{0 0 0 0 }k_{c1}J_{x}\text{ }k_{c2}J_{y}\right]  ^{T}$ and
$\bar{v}_{2}=\frac{v_{2}}{\left\Vert v_{2}\right\Vert }$, we have $\bar{A}%
^{T}\bar{v}_{2}=0$ and%
\[
\bar{v}_{2}^{T}\bar{B}\bar{u}=\frac{k_{c1}L+k_{c2}M}{\left\Vert v_{2}%
\right\Vert }.
\]
According to (\ref{case2-}),
\[
\underset{u\in\mathcal{U}_{\eta_{2}=0}^{0}}{\max\text{ }}\bar{v}_{2}^{T}%
\bar{B}\bar{u}=0.
\]
Thus, the system (\ref{linear_degraded}) is uncontrollable if $T\geq\frac
{18}{5}K$ according to \emph{Theorem 4}. Under hovering conditions, we have
$T=m_{a}g$. Thus, \emph{Proposition 1} is true.

\subsubsection{Proof of Proposition 2}

According to the proof of \emph{Proposition 1}, If $T<\frac{18}{5}K$, then
$c_{0}=\left[  \text{0 0}\right]  ^{T}$\ is an interior point of $E_{c}$.
According to \cite{Goodwin}, we cannot find a non-zero vector $c_{k}=\left[
k_{c1}\text{ }k_{c2}\right]  ^{T}$ satisfying $c_{k}^{T}c\leq0$ for all $c\in
E_{c}$. We will prove this by the proof of contradiction. Suppose that there
is a non-zero vector $\bar{v}_{2}=\left[  0\text{ }0\text{ }0\text{ }0\text{
}\bar{k}_{1}\text{ }\bar{k}_{2}\right]  ^{T}$ satisfying%
\[
\underset{u\in\mathcal{U}_{\eta_{2}=0}^{0}}{\max\text{ }}\bar{v}_{2}^{T}%
\bar{B}\bar{u}=0
\]
then we have%
\[
\bar{v}_{2}^{T}\bar{B}\bar{\delta}=\bar{k}_{1}L/J_{x}+\bar{k}_{2}M/J_{y}%
\leq0.
\]
Let $c_{k}=\left[  k_{c1}\text{ }k_{c2}\right]  ^{T}=\left[  \bar{k}_{1}%
/J_{x}\text{ }\bar{k}_{2}/J_{y}\right]  ^{T}$. Then we get
\[
c_{k}^{T}c=\bar{k}_{1}L/J_{x}+\bar{k}_{2}M/J_{y}\leq0
\]
and this contradicts with the fact that there is no non-zero vector $c_{k}$
satisfying $c_{k}^{T}c\leq0$ for all $c\in E_{c}$. Thus, the system
(\ref{linear_degraded}) is controllable if $T<\frac{18}{5}K$ according to
\emph{Theorem 4}. Under hovering conditions, we have $T=m_{a}g$. Thus,
\emph{Proposition 2} holds.

\end{document}